\documentclass[11pt]{article} 
\usepackage{amsmath} 
\usepackage{amssymb}
\usepackage{graphicx}
\usepackage{authblk}
\usepackage[usenames,dvipsnames]{color}
\usepackage[colorlinks=true,citecolor=blue,linkcolor=blue,urlcolor=blue]{hyperref}
\usepackage{cite}

\def\coeff#1#2{{\textstyle {\frac {#1}{#2}}}}
\def\Nf{{N_{\rm f}}}
\def\Nc{{N_{\rm c}}}
\def\k{{\bf k}}
\def\x{{\bf x}}
\def\q{{\mathfrak q}}
\def\w{{\mathfrak w}}
\def\s{{\mathfrak s}}
\def\E{{\mathfrak E}}

\begin{document}

\title{\LARGE\textbf{Instability in ${\cal N}=4$ supersymmetric Yang-Mills theory at finite density}}

\author[1]{Liam Gladden}
\author[2]{Victor Ivo}
\author[3]{Pavel Kovtun}
\author[1]{Andrei O. Starinets}
\affil[1]{\it The Rudolf Peierls Centre for Theoretical Physics, University of Oxford, Parks Road,  
Oxford, OX1 3PU, UK}
\affil[2]{\it Jadwin Hall, Princeton University, Princeton, NJ 08540, USA}
\affil[3]{\it Department of Physics \& Astronomy, University of Victoria, PO Box 1700 STN CSC, Victoria,
BC, V8W 2Y2, Canada}

\date{December, 2024}

\maketitle

\begin{abstract}
\noindent
Equilibrium states of ${\cal N}=4$ supersymmetric Yang-Mills theory can be characterized by the temperature and  three chemical potentials, corresponding to the ${\rm U}(1)^3$ subgroup of the $R$-symmetry group. We investigate the phase diagram of the theory at strong coupling  in the grand canonical ensemble in flat space, using its holographic description via the five-dimensional supergravity solution of Behrnd, Cveti\v{c},  and Sabra.  The bulk action includes the metric, three Abelian gauge fields, and two neutral scalar fields. The equilibrium state described by the charged black brane is always thermodynamically unstable at low temperature.  Relativistic hydrodynamics with multiple conserved charges predicts that thermodynamic instability is accompanied by a dynamical instability, with the eigenvalues and eigenvectors of the corresponding Hessian playing a key role in identifying the unstable modes.  We explicitly demonstrate this for three equal chemical potentials, finding unstable quasinormal modes that describe $R$-charge diffusion. Consequently, the low-temperature phase of ${\cal N}=4$ supersymmetric Yang-Mills theory with equal chemical potentials is not described by the AdS-Reissner-Nordstr\"om   black brane.
\end{abstract}

\section{Introduction}

Physical systems at finite temperature and density often undergo phase transformations as the temperature decreases. While the microscopic laws of many-body quantum physics are formulated in the language of quantum field theory, exact results concerning the phase structure of interacting field theories in 3+1 dimensions are limited. In this paper, we investigate the phase diagram of strongly interacting ${\cal N}=4$ supersymmetric $SU(\Nc)$ Yang-Mills theory (SYM) at finite temperature and density. This quantum field theory is interesting for a multitude of reasons, in particular forming the cornerstone of the holographic gauge-string duality~\cite{Maldacena:1997re, Gubser:1998bc, Witten:1998qj}. We will study this theory in flat space in the grand canonical ensemble. The theory features a global $SU(4)$ $R$-symmetry, with the corresponding maximal Abelian subalgebra $U(1)^3$. Consequently, equilibrium states can be analyzed at temperature $T$ and three chemical potentials~$\mu_a$, corresponding to the three $U(1)$ factors in $SU(4)$~\cite{Yamada:2006rx}. When both~$\Nc$ and the 't Hooft coupling of the SYM theory are large, holographic duality provides a description of the theory in terms of gravitational dynamics in asymptotically anti-de Sitter space (AdS)~\cite{Aharony:1999ti}. Here, we employ the five-dimensional STU black brane solution of Behrndt, Cveti\v{c}, and Sabra~\cite{Behrndt:1998jd}, which arises from a consistent truncation of ten-dimensional supergravity on $S^5$  \cite{Cvetic:1999xp,Cvetic:2000nc}. The solution includes the metric, three Abelian gauge fields for the three $U(1)$ factors in $SU(4)$, and two neutral scalar fields. When the chemical potentials are taken equal, $\mu_1 = \mu_2 = \mu_3$, the 
background reduces to the AdS$_5$-Reissner-Nordström (AdS$_5$-RN) black brane, with no scalar fields. We will focus on thermodynamic instabilities of ${\cal N}=4$ SYM and their (hydro)dynamic manifestations. 

Thermodynamic instabilities in the STU model have been known for some time \cite{Cvetic:1999ne, Cvetic:1999rb, Cai:1998ji, Gubser:2000ec, Gubser:2000mm, Son:2006em}, and the hydrodynamic instability in the solution with a single chemical potential was identified in ref.~\cite{Buchel:2010gd}. There is also a significant body of literature on STU black holes which are dual to ${\cal N}=4$ SYM on a three-sphere, as well as more general black holes involving charged scalars and their instabilities \cite{Chamblin:1999tk, Gubser:2008px, Dias:2009iu, Bhattacharyya:2010yg, Dias:2010eu,Dias:2010gk, Markeviciute:2016ivy, Yaffe:2017axl, Dias:2022eyq, Choi:2024xnv, Dias:2024edd}. In this work, we focus on ${\cal N}=4$ SYM in Minkowski space.

Using the standard criteria for thermodynamic stability~\cite{Callen} and analyzing the eigenvalues and eigenvectors of the relevant Hessian, we will show that five-dimensional STU black branes are thermodynamically unstable at low temperatures.  Our analysis is similar to the analysis of instabilities in charged AdS$_4$ black holes by Gubser and Mitra \cite{Gubser:2000ec, Gubser:2000mm}.  For STU black branes, hydrodynamics  predicts that thermodynamic instabilities lead to dynamic instabilities for the relevant fluctuation modes. As an illustration, we compute the quasinormal modes of STU black branes at $\mu_1 = \mu_2 = \mu_3$ and find that diffusive quasinormal modes move to the upper complex half-plane at the onset of 
thermodynamic instability, in agreement with relativistic hydrodynamics of theories with multiple charges which we develop in the paper.   In other words, the $R$-charge diffusion coefficient becomes negative at large $\mu/T$, resulting in a dynamical instability. 
The five-dimensional fluctuations responsible for this instability involve a combination of gauge fields and neutral scalars.

Before delving into the holographic description, we begin with a brief discussion of the relationship between thermodynamic and (hydro)dynamic instabilities.

\section{Thermodynamic and hydrodynamic (in)stability}
\label{sec:thermo-stability}

Consider a field theory with conserved charges $Q_a$ for a global symmetry $U(1)^\Nf$, with $a = 1, \dots ,\Nf$.  
The equation of state in the grand canonical ensemble is given by the pressure, $p = p(T,\mu_a)$, with the entropy density $s(T,\mu_a)$ and charge densities $n_a(T,\mu_a)$ determined by the first law of thermodynamics.
The equation of state in the microcanonical ensemble is given by $s = s(\epsilon, n_a)$, where $\epsilon$ is the (volume)  energy density.

In equilibrium, entropy is maximized, hence $s(\epsilon, n_a)$ is a concave function; correspondingly, the Hessian matrix of second derivatives of $s(\epsilon, n_a)$ must be negative definite. Alternatively, energy is minimized in equilibrium, hence $\epsilon(s, n_a)$ is a convex function; correspondingly, the Hessian matrix of second derivatives of $\epsilon(s, n_a)$ must be positive definite~\cite{Callen}. 
We will need the Hessian matrices $H^{\epsilon}_{ij} \equiv \partial^2 \epsilon/\partial y_i \partial y_j$, where $y_i = (s, n_a)$, and $H^p_{ij} \equiv \partial^2 p/\partial z_i \partial z_j$, where $z_i = (T,\mu_a)$. These can be rewritten as $H^\epsilon_{ij} = {\partial z_i}/{\partial y_j}$ and $H^p_{ij} = {\partial y_i}/{\partial z_j}$, hence $H^\epsilon = \left( H^p \right)^{-1}$. The condition of $H^\epsilon$ being positive definite is thus the same as the condition of $H^p$ being positive definite.
The latter implies that the leading principal minors of $H^p$ must be positive. In particular, in a thermodynamically stable system, the matrix of charge susceptibilities $\chi_{ab} \equiv (\partial n_a /\partial \mu_b)_T$ must be positive definite, as well as $T (\partial s/\partial T)_{\mu_a} >0$.  

When the conditions of thermodynamic stability are violated, the equilibrium state is unstable. By themselves, thermodynamic stability conditions do not say anything about the  dynamical realisation of the instability. However, for a  system which admits a hydrodynamic regime, thermodynamic and dynamic instabilities can be related. Indeed, hydrodynamics is a classical description of the slow, long-wavelength evolution of the densities of conserved charges (including $\epsilon$ and $n_a$) near thermal equilibrium. If a new equilibrium state with higher entropy can be achieved though a macroscopic change of conserved densities, the corresponding instability should be visible in hydrodynamics.

Consider small linearised fluctuations of energy density, momentum density, and charge densities proportional to $\exp(-i \omega t + i \k{\cdot}\x)$, where $\omega$ is the frequency and $\k$ is the wave vector. Hydrodynamic equations will give rise to dispersion relations $\omega = \omega(\k)$. Typical examples include sound, diffusion, and shear modes, which at small $\k$ behave as
\begin{align}
\label{eq:w-sound-diff}
  \omega_{\rm sound} = \pm v_s |\k| - i  \frac{\Gamma}{2} \k^2 + O(\k^3)\,,\ \ \ \ 
  \omega_{\rm diff} = - i D \k^2 + O(\k^4)\,,\ \ \ \ 
  \omega_{\rm shear} = - i \gamma \k^2\,.
\end{align}
Here $v_s$ is the speed of sound, $\Gamma$ is the sound damping coefficient, $D$ is the charge diffusion coefficient, and $\gamma$ is the diffusion coefficient for transverse momentum. A violation of thermodynamic stability conditions could conceivably give rise to ${\rm Im}\,v_s \neq 0$, or $\Gamma <0$, or $D<0$, or $\gamma<0$, signalling a hydrodynamic instability.  

For relativistic field theories, the corresponding dispersion relations follow from relativistic hydrodynamics~\cite{Kovtun:2012rj}. As a simple example, consider hydrodynamic fluctuations in a theory with one charge flavour, $\Nf=1$, about an equilibrium state with $\mu = 0$. The hydrodynamic dispersion relations for fluctuations of $\epsilon$ and $n$ are given by \eqref{eq:w-sound-diff}, with $v_s^2 = s/(T \partial s/\partial T)$, and $D = \sigma/(\partial n/\partial\mu)$, where $\sigma>0$ is the charge conductivity~\cite{Kovtun:2012rj}.  A violation of the thermodynamic stability condition $T \partial s/\partial T > 0$ leads to imaginary $v_s$, and hence to an unstable sound mode \cite{Buchel:2005nt}.  A violation of the thermodynamic stability condition $\partial n/\partial \mu > 0$ leads to negative $D$, and hence to an unstable diffusion mode. 
For fluctuations about a state with $\mu\neq0$, in a theory with $v_s^2>0$, a violation of thermodynamic stability condition for charge susceptibility can give rise to $D<0$, but not to $\Gamma<0$ \cite{Kovtun:2012rj}. Similarly, $\gamma = \eta/(\epsilon{+}p)$, where $\eta>0$ is the shear viscosity, and $\epsilon{+}p$ is the enthalpy density (momentum  susceptibility). A violation of the thermodynamic stability condition $\epsilon{+}p>0$ leads to negative $\gamma$, and to unstable shear modes.

\section{Thermodynamic instability of  the STU background}
\label{sec:thermo-inst}
Details of the five-dimensional STU background are well known (see Appendix~\ref{STU-BG}).  Standard black hole thermodynamics  (see e.g. \cite{Harmark:1999xt}, \cite{Son:2006em}) gives the equation of state
%
\begin{align}
\label{eq:eos-STU}
  \epsilon(s, n_1, n_2, n_3) = \frac{3}{2 (2\pi \Nc)^{2/3}} \, s^{4/3} \prod_{a=1}^3 \left(1+\frac{8\pi^2 n_a^2}{s^2} \right)^{1/3}\,,
\end{align}
where $s$ and $n_a$ are related to the parameters $T_0$, $\kappa_a$ of the gravitational background as 
\begin{align}
\label{eq:sn-kappa}
  s = \frac{\pi^2}{2} \Nc^2 T_0^3 \sqrt{(1{+}\kappa_1) (1{+} \kappa_2) (1 {+} \kappa_3)},\ \ \ \ 
  n_a = \frac{\pi}{8} \Nc^2 T_0^3 \sqrt{2\kappa_a}  \sqrt{(1{+}\kappa_1)(1{+}\kappa_2)(1{+}\kappa_3)} .
\end{align}
Given the equation of state \eqref{eq:eos-STU}, we can evaluate the Hessian $H^\epsilon$ and its leading principal minors.
\begin{figure}[t]
\centering
\includegraphics[width=0.45\textwidth]{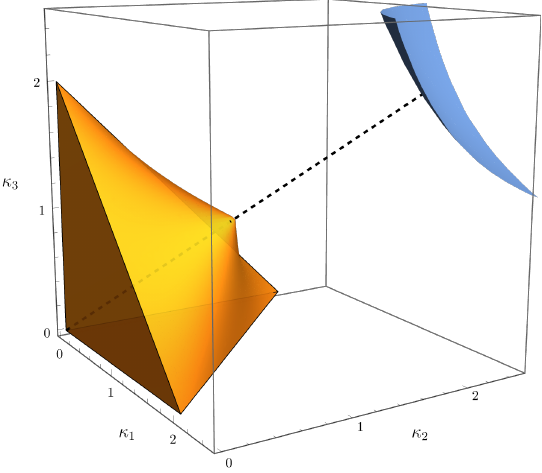}
\hspace{0.05\textwidth}
\includegraphics[width=0.45\textwidth]{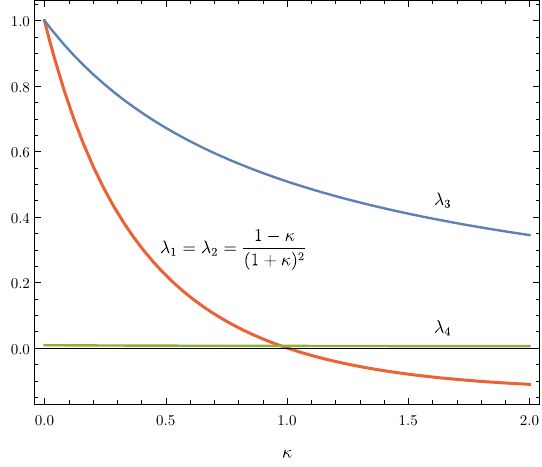}
\caption{
  Left: The phase diagram for the equation of state~\eqref{eq:eos-STU}, in the space of $\kappa_1$, $\kappa_2$, $\kappa_3$. The orange region connected to $\kappa_a = 0$ is the region of thermodynamic stability defined by eqs.~\eqref{eq:kappa-stability}, where the Hessian matrix of $\epsilon(s,n_a)$ is positive-definite. Outside the orange region, thermodynamic stability conditions are violated. The blue surface indicates $T=0$, according to eq.~\eqref{eq:TT0}. The dashed line marks $\kappa_1 = \kappa_2 = \kappa_3 = \kappa$; along this line, $T=0$ is achieved at $\kappa=2$. Right: The four eigenvalues $\lambda_i$ of the Hessian $h_{ij}$, as functions of $\kappa_1 = \kappa_2 = \kappa_3 = \kappa$. At $\kappa=1$, the two coinciding eigenvalues change sign, indicating a violation of thermodynamic stability conditions for $\kappa>1$.
}
\label{fig-kappa-space}
\end{figure}
Using eq.~\eqref{eq:sn-kappa}, thermodynamic stability conditions can be expressed in terms of $\kappa_a$ as 
\begin{subequations}
\label{eq:kappa-stability}
\begin{align}
  & 2 - \kappa_1 - \kappa_2 - \kappa_3 + \kappa_1 \kappa_2 \kappa_3 > 0 \,,\\
  & 3 - \kappa_1 - \kappa_2 - \kappa_3 - \kappa_1 \kappa_2 - \kappa_1 \kappa_3 - \kappa_2 \kappa_3 + 3 \kappa_1 \kappa_2 \kappa_3 >0\,,\\
  & 3 - \kappa_2 - \kappa_3 - \kappa_2 \kappa_3 > 0 \,, \\
  & 3 - \kappa_3 > 0 \,.
\end{align}
\end{subequations}
The region of thermodynamic stability defined by eqs.~\eqref{eq:kappa-stability} is shown in Fig.~\ref{fig-kappa-space}.  The region is concentrated  around $\kappa_a=0$ (equivalently, $\mu_a/T=0$). As the temperature is lowered, the  boundary of stability is inevitably crossed, and the low-temperature  state  is always unstable. 

We now focus on the state with $\kappa_1 = \kappa_2 = \kappa_3$. Going along the line $\kappa_a = \kappa$ in the phase diagram, the four eigenvalues $\lambda_i$ of the normalized Hessian $h_{ij} = \frac18 \Nc^2 T_0^2 H^\epsilon_{ij}$ are shown in Fig.~\ref{fig-kappa-space}. At $0<\kappa<1$, all four eigenvalues are positive, indicating a thermodynamically stable phase. At $1<\kappa<2$, two coinciding eigenvalues change sign, making the Hessian not positive-definite and violating thermodynamic stability conditions (note that the determinant of the Hessian remains non-negative). At equal $\kappa_a = \kappa$, the three chemical potentials for the three $U(1)$'s are also equal, $\mu_a = \mu$, with $\mu/2\pi T = \sqrt{2\kappa}/(2{-}\kappa)$.  The critical value $\kappa_c = 1$ thus corresponds to the critical value of the chemical potential $\mu_c/2\pi T = \sqrt{2}$. 

The eigenvalues of the susceptibility matrix $\chi_{ab}= \partial^2 p/\partial\mu_a \partial\mu_b$ comprise two coinciding eigenvalues $\chi_{11} {-} \chi_{12}$, and one eigenvalue $ \chi_{11} {+} 3 \chi_{12}$, with 
\begin{align}
\label{eq:chi-evals}
   \chi_{11} - \chi_{12} = \frac{\Nc^2 T_0^2}{8}  \frac{(1+\kappa)^2}{1-\kappa}\,,\ \ \ \ 
   \chi_{11} + 3 \chi_{12} = \frac{\Nc^2 T_0^2}{8}  \frac{(1+\kappa) (2+5\kappa)}{2+\kappa} \,.
\end{align}
The coinciding eigenvalues change sign at $\kappa{=}1$, indicating a violation of thermodynamic stability for $\kappa>1$. The eigenvalues diverge at $\kappa{=}1$, suggesting that the corresponding diffusion coefficients vanish at $\kappa=1$ (see Appendix~\ref{sec:dec}), and become negative for $\kappa>1$. We will see that this is indeed the case. 

In the space of $(s,n_i)$, the two eigenvectors of the Hessian $H^\epsilon$ which correspond to the two coinciding eigenvalues can be taken as two independent linear combinations of the three vectors $W_{12} = (0,1,-1,0)$, $W_{23} = (0,0,1,-1)$, $W_{31} = (0,-1,0,1)$, which satisfy $W_{12} + W_{23} + W_{31} = 0$. As the eigenvectors correspond to the directions in the $(s, n_i)$ space along which the energy can be lowered, one may guess that the fluctuations of $n_1 {-} n_2$, $n_2 {-} n_3$, and $n_3 {-} n_1$ will be unstable for $\mu > \mu_c$. Indeed, in the hydrodynamics of ${\cal N}=4$ SYM theory, charge density fluctuations $\delta (n_1 {-} n_2)$, $\delta(n_2 {-} n_3)$, and $\delta(n_3 {-} n_1)$ will have (the same) negative diffusion constant at $\mu>\mu_c$.  Alternatively, the eigenvectors corresponding to the two coincident eigenvalues of $H^\epsilon$ can be taken as two independent linear combinations of $V_{1} = (0,2,-1,-1)$, $V_{2} = (0,-1,2,-1)$, $V_{3} = (0,-1,-1,2)$, which again satisfy $V_{1} + V_{2} + V_{3} = 0$. Thus, one may guess that the fluctuations $3\phi_1 \equiv 2\delta n_1 {-} \delta n_2 {-} \delta n_3$, $3\phi_2 \equiv 2\delta n_2 {-} \delta n_1 {-} \delta n_3$, and $3\phi_3 \equiv 2\delta n_3 {-} \delta n_1 {-} \delta n_2$ will be unstable for $\mu>\mu_c$, and indeed they are. The fluctuations $\phi_a$ satisfy $\delta n_a = \coeff13 \delta(n_1 + n_2 + n_3) + \phi_a$, and thus may be interpreted as deviations of $\delta n_a$ from the ``average'' fluctuation $\coeff13 \delta(n_1 + n_2 + n_3)$.

\section{Hydrodynamic instability of the STU background}
\label{sec:hydro}
To uncover the dynamical nature of  the instability of the state with $\kappa_1 = \kappa_2 = \kappa_3$, we study small time-dependent perturbations of the STU black brane (see Appendix \ref{STU-BG}) with $\kappa_a = \kappa$. 
Rotation invariance implies that the fluctuations split into scalar channel, shear channel, and sound channel, just as they do for uncharged black branes~\cite{Kovtun:2005ev}. Hydrodynamic arguments, combined with the eigenvectors of the Hessian at  $\kappa_a = \kappa$, suggest that there will be a hydrodynamic instability in the sound channel, with a negative diffusion coefficient, for decoupled fluctuations $\delta n_a - \coeff13 \delta(n_1 + n_2 + n_3)$ at sufficiently small $T/\mu$. 

Working in the gauge $\delta A^a_u = 0$, we take $\delta A^i_\mu = \sqrt{2}\, \pi T_0 (1+\kappa)^{3/2} a^i_\mu(u) e^{-i\omega t + i k z}$, and define $\w_0 \equiv \omega/2\pi T_0$, $\q_0 \equiv k/2\pi T_0$. The corresponding ``electric fields'' are $E^i_z \equiv \w_0 a^i_z + \q_0 a^i_t$, and the scalar fluctuations are $\delta H^i = s^i(u) e^{-i\omega t + i k z}$. We next define
\begin{align}
  \E^a_z \equiv  E^a_z - \coeff13(E^1_z + E^2_z + E^3_z) \,,\ \ \ \ 
  \s^a \equiv s^a -\coeff13 (s^1 + s^2 + s^3) \,.
\end{align}
Using the equations of motion which follow from \eqref{eq:L5}, one can show that the perturbations $\E^a_z(u)$ and $\s^a(u)$ indeed decouple for each flavour ``$a$'', and satisfy the following equations (we omit index ``$a$''):
\begin{align}
   \E_z{}'' 
   &+ \frac{1}{G} \Biggl[ \w_0^2 (1+\kappa u)^3\left( \frac{f'}{f} - \frac{\kappa}{1+\kappa u}\right) - \frac{2\kappa  \mathfrak{q}_0^2f}{1+\kappa u} \Biggr]  \mathfrak{E}_z{}'  
   + \frac{G}{uf^2}  \mathfrak{E}_z 
   \nonumber \\
   &+ \frac{2\sqrt{\kappa}\, \mathfrak{q}_0}{(1+\kappa u)^3} \, \mathfrak{s}{}'  
    + \frac{2\sqrt{\kappa}\, \mathfrak{q}_0}{G}  \Biggl[\mathfrak{w}_0^2\left(\frac{f'}{f} -\frac{4\kappa}{1+\kappa u}\right) + \frac{\kappa\mathfrak{q}_0^2  f}{(1+\kappa u)^4}\Biggr]\mathfrak{s}  = 0\,,
\label{eq:fluct-01}
\end{align}
\begin{align}
  \mathfrak{s}{}'' 
  &+ \left( \frac{f'}{f} - \frac{1+3\kappa u}{u(1+\kappa u)}\right) \mathfrak{s}{}' 
+\Biggl[  \frac{G}{uf^2} + \frac{1+\kappa u}{u^2f} + \frac{2\kappa (1+\kappa)^3u}{(1+\kappa u)^2f} 
  - \frac{\kappa}{1+\kappa u}\left(\frac{f'}{f} - \frac{1+3\kappa u}{u(1+\kappa u)}\right) 
\nonumber\\
  & -\frac{4\kappa (1+\kappa )^3(1+\kappa u)u \mathfrak{w}_0^2}{f G}\Biggr] \mathfrak{s}
  -  \frac{2\sqrt{\kappa} (1+\kappa)^3(1+\kappa u)u  \mathfrak{q}_0}{G} \, \mathfrak{E}_z{}' = 0 \,.
\label{eq:fluct-02}
\end{align}
Here,  $f(u)\equiv (1+\kappa u )^3 - u^2 (1+\kappa)^3$ and $G(u) \equiv (1+\kappa u)^3\mathfrak{w}_0^2 - \mathfrak{q}_0^2f$.
Solutions to eqs.~\eqref{eq:fluct-01} and \eqref{eq:fluct-02} which are normalizable at $u=0$ and infalling at $u=1$ represent quasinormal modes for a subset of sound channel fluctuations. Quasinormal mode solutions will  exist if the parameters $\w_0$ and $\q_0$ are related. In the following, we will use the normalized $\w \equiv \omega/2\pi T$, $\q \equiv k/2\pi T$, where $T$ is the Hawking  temperature \eqref{eq:TT0}.

At $u=1$, the indicial exponents for both $\E_z$ and $\s$ are $\pm i \w/2$. Following the recipe of  ref.~\cite{Son:2002sd}, we choose $-i\w/2$ for both $\E_z$ and $\s$, corresponding to infalling waves at the horizon. At $u=0$, the indicial exponents for $\E_z$ are $0$ and $1$; we choose the latter for the normalizable solution, just like in the case of uncharged branes. The indicial exponents at $u=0$ for $\s$ are two coincident exponents equal to $1$. Scalar fluctuations thus behave near $u=0$ as $ \s = {\cal A} u \log{u}+\cdots +{\cal B} u+\cdots$,  and the standard holographic recipe to obtain the poles of the dual retarded correlation functions~\cite{Kovtun:2005ev} requires setting ${\cal A}=0$. 

\begin{figure}[t]
\centering
\includegraphics[width=0.45\textwidth]{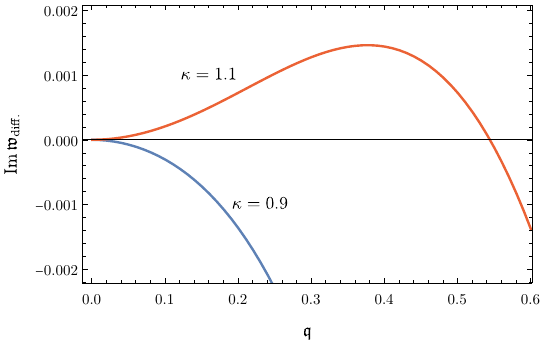}
\hspace{0.05\textwidth}
\includegraphics[width=0.45\textwidth]{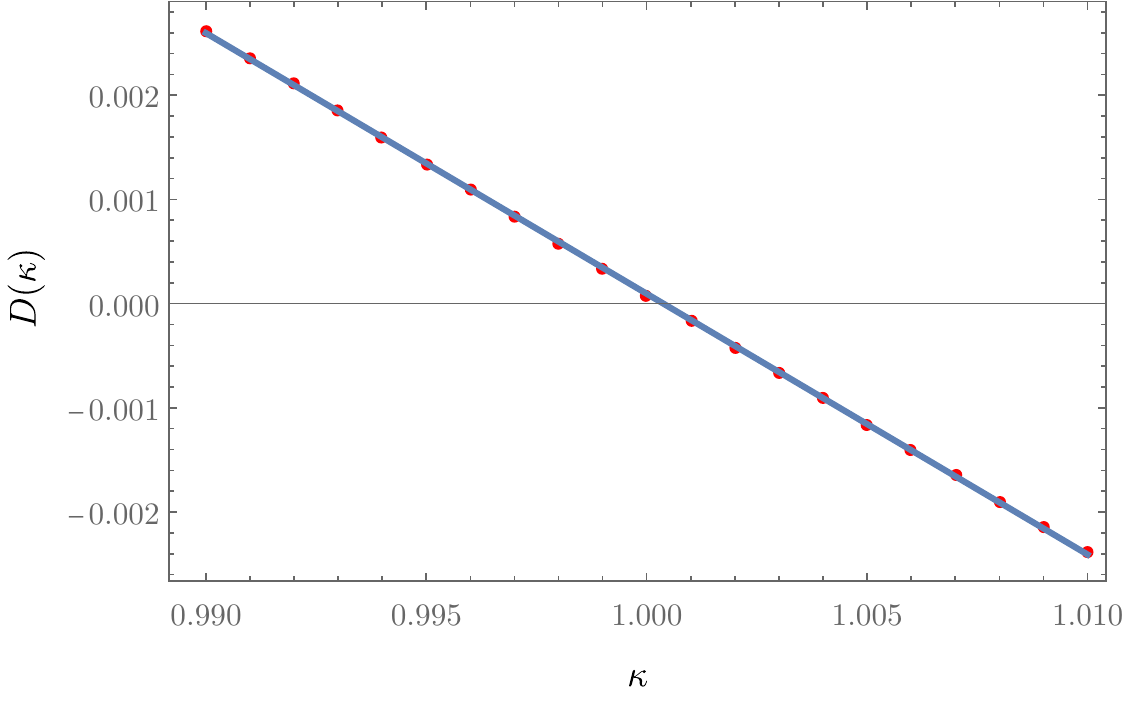}
\caption{
  Left: Imaginary part of the diffusive quasinormal frequency $\w_{\rm diff}$, plotted as a function of $\q \equiv k/2\pi T$, at $\kappa=0.9$ and $\kappa=1.1$. For $\kappa>1$, the diffusive mode is unstable at small $\q$, but becomes stable again at large $\q$. Right: The diffusion coefficient $D(\kappa)$ near $\kappa=1$. 
}
\label{fig:unstable-diffusion}
\end{figure}

Given the above boundary conditions, equations \eqref{eq:fluct-01} and \eqref{eq:fluct-02} can be solved numerically. At each fixed $\q$, there is an infinite set of quasinormal frequencies $\w_n(\q)$. As expected from hydrodynamics, this set contains a diffusive mode with $\w_{\rm diff}(\q) \sim -i D(\kappa) \q^2$ as $\q\to0$, with a real diffusion coefficient $D(\kappa)$. Our numerical results for the diffusion mode are shown in Fig.~\ref{fig:unstable-diffusion}.  At $\kappa<1$, $\w_{\rm diff}(\q)$ is in the lower complex half-plane for all $\q$, however, at $\kappa>1$, $\w_{\rm diff}(\q)$ crosses into the upper half-plane at small $\q$, indicating a hydrodynamic instability. 

The hydrodynamic prediction near $\kappa=1$ gives $D = (\sigma_{11} {-} \sigma_{12}) (1 {-} \kappa)/ {\Nc^2 T^2}$, where $\sigma_{ab}$ is the (positive-definite) conductivity matrix. Indeed, our numerical results give a linear dependence of $D$ on $(1{-}\kappa)$, see Fig.~\ref{fig:unstable-diffusion}, right. This is consistent with $\sigma_{11} - \sigma_{12}$ staying finite and non-zero as $\kappa\to1$ (as happens for other transport coefficients~\cite{Son:2006em}), and the dynamic instability arising purely as a consequence of thermodynamic instability. 

\section{Discussion}
Thermodynamic and hydrodynamic instabilities are not independent---we have illustrated their relationship using the STU solution of five-dimensional supergravity as an example. The eigenvalues and eigenvectors of the thermodynamic Hessian matrix are instrumental in identifying unstable modes in the hydrodynamic regime. We have found that STU black branes are both thermodynamically and hydrodynamically unstable at low temperatures. At large $T/\mu_a$, these black branes describe stable equilibrium states of the ${\cal N}=4$ SYM theory in flat space. Thus, holography implies that one cannot cool down ${\cal N}=4$ SYM in flat space from high to low $T$ without encountering instabilities (or, perhaps, phase transitions which may set in before the instability). 

The low-temperature instability discussed here persists even if all $\mu_a$ are equal, in which case the equilibrium background is described by the Reissner-Nordstr\"om black brane. This instability is invisible in a model containing only the metric and a single $U(1)$ gauge field $A_\mu$. However, when the AdS$_5$-RN background is embedded in the STU model---which includes three $U(1)$ gauge fields for $U(1)^3 \subset SU(4)$ (along with neutral scalars)---thermodynamic instability develops at low $T$. This thermodynamic instability is accompanied by a hydrodynamic instability. In other words, the AdS$_5$-RN background with $A_\mu^1 = A_\mu^2 = A_\mu^3$ is hydrodynamically unstable to fluctuations of $A_\mu^a - A_\mu^{b \neq a}$ (coupled to neutral scalars) at small $T/\mu$.

This situation is somewhat analogous to ``holographic superconductor'' models~\cite{Gubser:2008px, Herzog:2009xv}, where charged scalar fields in AdS lead to low-temperature instabilities that are invisible if charged scalars are neglected. However, the STU model does not contain charged scalar fields, and its low-$T$ instability is of a different nature. Importantly, the STU background model is not a ``bottom-up'' construction but rather a consistent truncation of IIB supergravity that describes ${\cal N}=4$ SYM theory. Consequently, the low-temperature phase of strongly coupled ${\cal N}=4$ SYM theory with $\mu_1 = \mu_2 = \mu_3$ is not described by the AdS$_5$-RN background.

Our primary focus in this paper was on the state with $\mu_1 = \mu_2 = \mu_3$. States with other ratios among the three $\mu_a$'s are also thermodynamically unstable at low temperatures, and we expect them to be hydrodynamically unstable as well. For states where only one of the three $\mu_a$'s is non-zero, thermodynamic stability arguments combined with the form of hydrodynamic dispersion relations~\cite{Kovtun:2012rj} imply that the charge diffusion mode is unstable at low temperatures. This instability was indeed observed in holography~\cite{Buchel:2010gd}. 

The instability discussed in this paper suggests a phase transition or the end of the phase diagram in the grand canonical ensemble as ${\cal N}=4$ SYM is cooled down to low~$T$.  Understanding the  low-$T$ behaviour of   ${\cal N}=4$ SYM  at finite density thus requires more work. Additional instabilities may arise when fluctuations of additional fields (whose STU background values vanish) are considered, either in five dimensions \cite{Dias:2022eyq} or ten dimensions \cite{Choi:2024xnv}. Clarifying the full phase structure of ${\cal N}=4$ SYM at finite density remains a desirable goal in both field theory and dual gravity/string theory.

Finally, we note that the relationship between thermodynamic and hydrodynamic stability is quite general, see e.g.~\cite{Gubser:2000ec, Gubser:2000mm, Buchel:2005nt, Friess:2005zp, Buchel:2010wk, Buchel:2011ra, Buchel:2020jfs, Gouteraux:2024adm}. 
We leave a more detailed analysis of thermodynamic and dynamic instabilities in STU black branes and other systems for future work~\cite{n4stabx}.

\section*{Acknowledgements}

We would like to thank Jay Armas, Alex Buchel, Oscar Dias and Laurence G. Yaffe for correspondence and Chris Herzog for discussions. PK would like to thank the Rudolf Peierls Centre for Theoretical Physics at Oxford for hospitality. The work of PK was supported in part by the NSERC of Canada.

\appendix

\section{The STU background}
\label{STU-BG}

As a minimal gravity background dual to ${\cal N}=4$ SYM theory at finite density of $R$-charge,  we will use the  STU solution of ref.~\cite{Behrndt:1998jd} which, in addition to the three gauge fields $A_\mu^a$ ($a=1,2,3$) also contains real positive scalar fields $X^a$, constrained by $X^1 X^2 X^3 =1$.
The three $X^a$ can be written as exponentials of two real scalar fields $\varphi_{1,2}$.
The five-dimensional action is $S = \frac{1}{16\pi G_5} \int d^5 x \sqrt{-g}\, {\cal L}$, where the Newton's constant $G_5$ is related to the AdS radius $L$ and the number of colours in the SYM theory as $G_5 = \pi L^3/2 \Nc^{2}$. The Lagrangian is
\begin{align}
\label{eq:L5}
  {\cal L} = R - \frac{L^2}{4} G_{\!ab} F_{\mu\nu}^a F_{\mu\nu}^b  + \frac{c_{abc}}{24} \frac{L^3}{2^{3/2}} \epsilon^{\mu\nu\rho\sigma\lambda} F_{\mu\nu}^a F_{\rho\sigma}^b A_\lambda^c - G_{\!ab} g^{\mu\nu} \partial_\mu X^a \partial_\nu X^b + \frac{4}{L^2} \sum_{a=1}^3 \frac{1}{X^a}\,,
\end{align}
where $g_{\mu\nu}$ is the metric, $R$ is the Ricci scalar, $F_{\mu\nu}^a$ is the field strength for the gauge field $A_\mu^a$, and $G_{ab} = \frac12 {\rm diag}((X^1)^{-2}, (X^2)^{-2}, (X^3)^{-2})$. The coefficients $c_{abc}$ are completely symmetric,  their values can be chosen so that $c_{abc}=1$ for $abc=123$ and permutations, with the rest of $c_{abc}$ zero. In the AdS/CFT correspondence, the coefficients $c_{abc}$ are proportional to the anomaly coefficients $d_{abc}$~\cite{Witten:1998qj, Freedman:1998tz}; the choice $c_{abc}=1$ translates to one's choice of the generators of the $U(1)^3 \subset SU(4)$. When all $X^a=1$,   the potential term for $X^a$ becomes $-2\Lambda$, where $\Lambda = -6/L^2$ is the cosmological constant expressed in terms of the AdS radius.
We will also make use of the scalar fields $H^a$ which are related to $X^a$ through $X^a = H^{1/3}/H^a$, with $H\equiv H^1 H^2 H^3$.

The equations of motion for the action \eqref{eq:L5} admit non-extremal three-brane solutions which depend on four parameters which indirectly determine the temperature $T$ and the three chemical potentials $\mu_a$ of the dual SYM theory. We will call the four parameters of the black brane solution $T_0$ and $\kappa_a$; their relations to $T$ and $\mu_a$ of the SYM theory appear below. We use the coordinates $t, x, y, z, u$, where $u=1$ is the black brane horizon, and $u=0$ is the asymptotic infinity. The scalar fields $H^a$ of the black brane background are 
\begin{align}
\label{eq:H-bg}
  H^a(u) = 1 + \kappa_a u\,,
\end{align}
hence $H(u) = (1 {+} \kappa_1 u) (1 {+} \kappa_2 u) (1 {+} \kappa_3 u)$. The gauge fields of the black brane background are
\begin{align}
\label{eq:A-bg}
  A_\mu^a(u) = \delta_\mu^t \left( \frac{1}{1+\kappa_a} - \frac{u}{H^a(u)} \right) \pi T_0 \sqrt{2\kappa_a} \sqrt{(1+\kappa_1)(1+\kappa_2)(1+\kappa_3)}\,,
\end{align}
where the integration constant has been chosen so that $A_t^a$ vanishes at the horizon $u{=}1$.
Finally, the black brane metric is
\begin{align}
\label{eq:STU-metric}
  ds^2 = - H^{-2/3}{(\pi T_0 L)^2 \over u}\,f \, dt^2 
+   H^{1/3}{(\pi T_0 L)^2 \over u}\, \left( dx^2 + dy^2 + dz^2\right)
+  H^{1/3}{L^2 \over 4 f u^2} du^2\,,
\end{align}
where $f(u) =  H(u) - u^2  H(1)$. The Hawking temperature  of the background is
\begin{align}
\label{eq:TT0}
  T = \frac{2+\kappa_1 + \kappa_2 + \kappa_3 - \kappa_1 \kappa_2 \kappa_3}{2\sqrt{(1+\kappa_1) (1+\kappa_2) (1+\kappa_3)}} T_0 \,.
\end{align}
The chemical potential is determined by the asymptotic value of the gauge field, $\mu_a = A_t^a(u{=}0)$,
\begin{align}
  \mu_a = \pi T_0 \frac{\sqrt{2\kappa_a}}{1+\kappa_a} \sqrt{(1+\kappa_1)(1+\kappa_2)(1+\kappa_3)}\,.
\end{align}
Standard black hole thermodynamics \cite{Son:2006em} gives the equation of state \eqref{eq:eos-STU} with the parameters \eqref{eq:sn-kappa}.
Consider now the state with three equal $\kappa_a = \kappa$. Such a state also has equal $\mu_a = \mu$ and equal $H^a$.  
Correspondingly, $X^a = 1$, the scalar fields $\varphi_{1,2}$ vanish, and the background is given by the AdS Reissner-Nordstr\"om (RN) black brane, which therefore describes the equilibrium state of ${\cal N}=4$ SYM theory at non-zero temperature $T$ and equal non-zero chemical potentials $\mu_a = \mu$ for the three $U(1)$ R-charges. This AdS$_5$-RN background is only stable at high~$T$.

\section{Diffusion in a state with equal charges}
\label{sec:dec}
To clarify thermodynamic stability conditions in states with non-zero $\mu_a$,  consider hydrodynamics of a relativistic system with $\Nf$ conserved $U(1)$ charges. In the conventions of Landau and Lifshitz~\cite{LL6}, the constitutive relations for the $U(1)$ currents are
\begin{align}
\label{eq:Ja-const-rel}
  & J^\mu_a = n_a u^\mu - \sigma_{ab} T \Delta^{\mu\nu} \partial_\nu \gamma_b + O(\partial^2) \,.
\end{align}
Here $u^\mu$ is the fluid velocity (normalized such that $u^2=-1$), $\Delta^{\mu\nu} \equiv g^{\mu\nu} + u^\mu u^\nu$ is the projector onto the space orthogonal to $u$, and $\gamma_a\equiv \mu_a/T$.  The charge densities $n_a(T, \gamma_1, \gamma_2, \dots)$ are given by the equilibrium equation of state. The coefficient $\sigma_{ab}$ is the matrix of charge conductivities.  We assume unbroken charge conjugation symmetry for each flavour, and further assume that the microscopic theory respects time-reversal symmetry, in which case the conductivity matrix $\sigma_{ab}$ will be symmetric by Onsager relations. The divergence of the entropy current has an additive contribution $\sigma_{ab} \Delta^{\mu\nu} \partial_\mu \gamma_a \partial_\nu \gamma_b$. Hence, the matrix $\sigma_{ab}$ must also be positive semi-definite, in order to ensure non-negative entropy production out of equilibrium~\cite{LL6}. The requirement that the hydrodynamic entropy current has a non-negative divergence places positivity constraints on transport coefficients such as $\sigma_{ab}$. However, it does not constrain the equation of state: that is done by thermodynamic stability conditions. As the example of diffusion in the main text illustrates, demanding $\sigma>0$ does not guarantee that the equilibrium state will be stable: one must also ensure that the susceptibility $\partial n/\partial\mu$ is positive. 

Consider now small fluctuations about an equilibrium state of the fluid at rest with constant temperature and chemical potentials. Namely, we take $u^\lambda = (1, {\bf 0}) + \delta u^\lambda$, and similarly for $T$ and $\mu_a$. For linearized fluctuations, current conservation of eq.~\eqref{eq:Ja-const-rel} then gives: 
\begin{align}
\label{eq:J-linear-2}
     \left( \frac{\partial n_a}{\partial T}\right)_{\!\!\gamma}  \partial_t \delta T +  n_a \, \partial_i \delta u^i + T \, {\cal D}_{ab} \delta \gamma_b  = 0 \,,
\end{align}
where ${\cal D}_{ab} \equiv \chi_{ab} \partial_t - \sigma_{ab} \partial_i \partial^i$ is the diffusion operator, ${\cal D}_{ab} = {\cal D}_{ba}$, and all quantities which multiply the derivatives of $\delta T$, $\delta u^i$, $\delta\gamma$ are evaluated in equilibrium. 
In writing down this equation, we used $(\partial n_a/\partial \gamma_b)_T = T \chi_{ab}$ where $\chi_{ab} \equiv (\partial n_a/\partial\mu_b)_T$ is the matrix of charge susceptibilities.  Eq.~\eqref{eq:J-linear-2} implies is that in states with non-zero charge densities in equilibrium, the fluctuations of temperature, velocity, and the chemical potentials are in general coupled. However, there can be situations when some of the fluctuations $\delta \gamma_a$ decouple from $\delta T$ and $\delta u^i$. 

As an example of such decoupling, consider an equilibrium state with all chemical potentials equal, $\mu_a = \mu$. Next, let us assume that there is a symmetry which ensures that all charge densities in this state are also equal, $n_a(T, \mu_1 {=} \mu, \mu_2 {=} \mu, \dots) = n(T,\mu)$. Taking pair-wise differences of the individual equations in~\eqref{eq:J-linear-2}, one obtains $\Nf(\Nf-1)/2$ equations for $\delta\gamma_a$ which do not contain $\delta T$ and $\delta u^i$,
\begin{align}
\label{eq:gg1}
  \left({\cal D}_{ac} - {\cal D}_{bc} \right) \delta \gamma_c = 0 \,.
\end{align}
Next, suppose that the symmetry among the charges is such that when all $\mu_a = \mu$, all diagonal components of ${\cal D}_{ab}$ are equal to each other, and all off-diagonal components of ${\cal D}_{ab}$ are equal to each other (the $SU(4)$ R-symmetry of SYM will achieve this~\cite{n4stabx}). Taking $\Nf=3$, we have
\begin{align}
   \left( {\cal D}_{11} {-} {\cal D}_{12} \right) (\delta\gamma_1 {-} \delta\gamma_2) = 0,\ \ \ \ 
   \left( {\cal D}_{11} {-} {\cal D}_{12} \right) (\delta\gamma_2 {-} \delta\gamma_3) = 0,\ \ \ \ 
   \left( {\cal D}_{11} {-} {\cal D}_{12} \right) (\delta\gamma_1 {-} \delta\gamma_3) = 0.
\end{align}
Of the above three equations, only two are independent. We thus have two diffusive modes, say the fluctuations of $\gamma_1 {-} \gamma_2$ and $\gamma_2{-}\gamma_3$, each with the diffusion coefficient 
\begin{align}
\label{eq:d12}
  D = \frac{\sigma_{11} - \sigma_{12}}{\chi_{11} - \chi_{12}} \,.
\end{align}
Recall that for the matrices $\chi_{ab}$ and $\sigma_{ab}$ in a state with equal chemical potentials, $\chi_{11} - \chi_{12}$ is an eigenvalue of $\chi_{ab}$, and $\sigma_{11} - \sigma_{12}$ is an eigenvalue of $\sigma_{ab}$. 
Just like for a single $U(1)$ charge, positive hydrodynamic entropy production ($\sigma_{11}>\sigma_{12}$) together with a thermodynamically unstable equation of state ($\chi_{11} < \chi_{12}$) give rise to $D<0$, hence to  unstable diffusive modes.

We have phrased the above discussion in terms of $\delta \gamma_a$, but it can also be phrased in terms of $\delta n_a = \chi_{ab} \delta\mu_b + (\partial n_a/\partial T)_\mu \delta T$. Assuming that in equilibrium $n_a = n$, and that $\chi_{ab}$ has all diagonal elements equal, and all off-diagonal elements equal, we have $\delta n_1 - \delta n_2 = (\chi_{11} - \chi_{12}) (\delta \mu_1 - \delta \mu_2)$. At $\mu_a = \mu$, one has $\delta \gamma_1 - \delta\gamma_2 = (\delta \mu_1 - \delta\mu_2)/T$, and therefore the fluctuations of $n_1 {-} n_2$ are also diffusive, with diffusion coefficient \eqref{eq:d12}.

\bibliographystyle{JHEP}
\bibliography{instability-short}{}

\end{document}